\documentstyle[a4,11pt,epsf]{article}
\begin{document}

\textwidth    170mm
\textheight   250mm
\pagestyle{empty}
\hoffset 0 mm
\topmargin -1.5 cm
\baselineskip=16pt

\begin{flushright}
\begin{tabular}{l}
FTUAM-97/5 \\
hep-ph/9705285
\end{tabular}
\end{flushright}

\vskip 1cm

\vspace*{1.5cm}
 
{\vbox{\centerline{{\Large{\bf HEAVY HIGGS: IS THE GAUGE SYMMETRY}}}}}
 
\vskip12pt
 
{\vbox{\centerline{{\Large{\bf RESTORED IN THE EARLY UNIVERSE?}}}}}
 
\vskip 1 cm
\centerline{{ S. Vargas-Castrill\'on}}
\centerline{{\it Dept. de F\'{\i}sica Te\'orica C-XI}}\vskip2pt
\centerline{{\it Universidad Aut\'onoma de Madrid}}\vskip2pt
\centerline{{\it E-28049 Madrid, Spain}}

\vskip 2 cm

\begin{abstract}

The possibility of symmetry non-restoration at high temperature is explored in a strongly  
interacting Higgs system described by an effective Chiral Lagrangian. 
Despite a na\"{\i}ve perturbative hint of symmetry 
non-restoration when the temperature is increased, non-perturbative 
methods point towards symmetry restoration.

\end{abstract}

\vskip 1cm 

Contribution to the XXXIInd {\it Rencontres de Moriond}, ``Electroweak 
Interactions and Unified Theories'', Les Arcs, France, March 1997.

\vfill

\newpage

This talk is based on  work done in collaboration with M.B. Gavela, 
M.J. Herrero, O. P\`{e}ne and N. Rius.

Common intuition tells that, due to thermal agitation, symmetries are 
restored when a system 
is heated up. This can be physically understood as due to 
thermal excitations, which make possible for the system to cross the barrier surrounding 
the broken minimum and feel the full symmetry of the theory. This type of behaviour can be 
parametrized by means of a temperature dependent effective potential 
which  is 
computable order by order in perturbation theory and displays the whole 
symmetry of the theory.

The behaviour at high temperature of systems with symmetries was formally 
studied in the 70's by 
applying the techniques of Finite Temperature Field Theory. Symmetry 
restoration 
was generally found for both global [1] and gauge symmetries [2, 3]. 
This is, for example, the situation concerning the Standard Model with one 
Higgs doublet.   

As counterintuitive as it may seem, one wonders whether it is possible 
to find the opposite 
behaviour: a broken symmetry which gets more broken at high temperature, 
or even the more 
radical situation where an unbroken symmetry gets broken at high 
temperature [3, 4]. Weinberg 
[3] has already pointed out such a behaviour in a two Higgs multiplet model. 
He also mentions a beautiful experimental example borrowed from Solid State 
Physics: a ferroelectric crystal, the 
Rochelle salt, which presents a 
certain temperature region where an unbroken symmetry gets broken at high temperature, 
though as the temperature is further increased the symmetry becomes 
eventually restored [5]. 

Most extensions of the Standard Cosmological Big Bang Model are still 
afflicted from problems 
concerning so-called topological defects (domain walls, monopoles) which are 
supposed to be produced during phase transitions. Symmetry 
non-restoration (SNR) could provide a radical 
cure to this problem [6], avoiding phase transitions at all and therefore 
the generation of topological deffects. 

At present, there are two main avenues to explore physics beyond the 
Standard Model: theories in which the Higgs particle is a 
fundamental one, 
supersymmetry being its most relevant example and those for which the 
Higgs it is not, and a non-perturbative regime is 
appropriate. It is known
that internal symmetries are always restored for renormalizable supersymmetric 
theories [7]. For the latter, a recent analysis 
for systems involving non-vanishing background charges shows that SNR 
could  be possible [8]. The case of non-renormalizable theories 
has been discussed in [9].

To illustrate the
possibility of SNR in gauge theories, we use a simple model 
given by a two Higgs doublet model with a $SU(2)\times U(1)$ local symmetry. 
A general 
 renormalizable scalar potential having this symmetry is written as
\begin{eqnarray}
V(\phi,\psi) &=& -m_{1}^{2}\phi^{+}\phi-m_{2}^{2}\psi^{+}\psi+\lambda_{1}(\phi^{+}\phi)^{2} 
+\lambda_{2}(\psi^{+}\psi)^{2}\nonumber \\
&+&\lambda_{3}(\phi^{+}\phi)(\psi^{+}\psi)+\lambda_{4}
|\phi^{+}\psi|^2 + \ldots
\end{eqnarray}

Where $\phi$, $\psi$ are scalar fields and dots represent terms that are 
not relevant for this 
problem. Temperature corrections to the potential are readily obtained at 
one-loop level in the high temperature limit ($T \gg m_i$) [10] 
with the result, \begin{eqnarray}
V^{T}(\phi,\psi)&=&\frac{T^2}{24}[(6\lambda_1+2\lambda_2+\lambda_4+\frac{9}{4}g^2+
\frac{3}{4} g'^{2}\nonumber \\
&+& 3h_{b}^{2}+h_{\tau}^{2})v_{1}^{2}+(6\lambda_{2}+2\lambda_{3}+\lambda_{4}+
\frac{9}{4}g^{2}+\frac{3}{4}g'^{2}+3 h_{t}^{2})v_{2}^{2}].
\label{v}
\end{eqnarray}
Once the boundedness conditions are satisfied, the contribution from both 
fermions and gauge bosons is positive, tending
to restore the symmetry. The only possible source of SNR is therefore
provided by the scalar sector of the theory. Studying (\ref{v}) we have found that
SNR is possible whenever the following condition is satisfied,
\begin{equation}
6 \lambda_{1} + 2 \lambda_{3} + \lambda_{4} + \frac{9}{4} g^{2} + \frac{3}{4} g'^{2} 
+ 3 h_{b}^{2} + h_{\tau}^{2} < 0,
\end{equation}

that is, since $\frac{9}{4}g^2$ is already of order one, some of the {\it 
scalar couplings}
have to be negative and {\it greater than one in absolute value}, going
out of the perturbative regime, where the approach cannot be trusted any 
longer. To bypass this
problem we explore this non-perturbative region using a Chiral Lagrangian 
approach for a theory with a strongly interacting Higgs sector.

The ungauged Chiral Lagrangian can be written,
\begin{equation}
{\cal L}_{ChL} = \frac{v^{2}}{4} {\rm Tr} [\partial_{\mu} U^{\dag} 
\partial^{\mu} U], \label{chl}
\end{equation}
where $U = \exp{\frac{i \vec{\tau} {\vec\pi}}{v}}$. It is easy to 
convince oneself that
(\ref{chl}) is $SU(2)_{L} \times SU(2)_{R}$ invariant. This kind of Lagrangians are 
frequently used to describe pion interactions in QCD at low energies ($E 
< 1$ GeV) 
[11]. One of the most beautiful features of this approach is its 
universality, 
in the sense that the same Lagrangian can be used to describe all 
theories that show
the same pattern of symmetry breaking (e.g. $SU(2)_{L} \times 
SU(2)_{R}\rightarrow 
SU(2)_{L + R}$), the only change being the different values of the only 
dimensional parameter of the Lagrangian, $v$, which  is about $100$ MeV 
for QCD and $250$ GeV for the electroweak theory.

 Figure 1 
shows a clear tendency towards chiral symmetry restoration 
confirming previous results [12], 
although it is important to keep in mind that an exact number for the critical temperature 
cannot be obtained with this technique which is only valid for $T < 4 \pi v$. 

\begin{figure}
\begin{center}
\makebox[5cm]{
\epsfysize 5cm
\epsfbox[50 50 410 302]{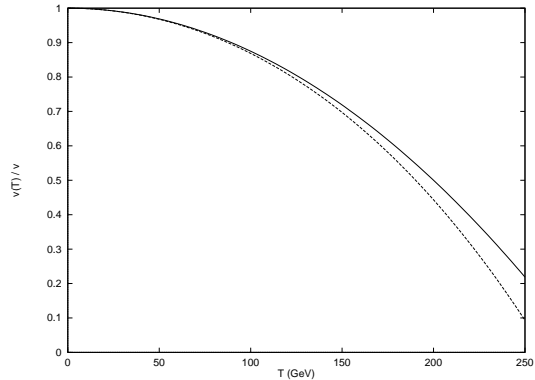}
}
\caption{Vacuum expectation value as a function of temperature at one 
loop (continuous line) and at two loops (dashed line).} \label{figura}
\end{center}
\end{figure}

Now we would like to construct, starting from (\ref{chl}), a locally 
(gauge) $SU(2)_L$ invariant chiral lagrangian, 
\begin{equation}
{\cal L}_{GChL} = \frac{v^2}{4}{\rm Tr} [D_{\mu}U^{\dag}D^{\mu}U]+{\cal 
L}_{YM}+{\cal L}_{GF} +{\cal L}_{FP}, 
\label{gcl}
\end{equation}
where the covariant derivative is given by $D_{\mu} U=\partial_{\mu}U+
i\frac{g}{2}(\vec{W}_{\mu} \vec{\tau}) U$. ${\cal L}_{YM}$ is the pure 
Yang-Mills 
Lagrangian, ${\cal L}_{GF}$ and ${\cal L}_{FP}$ the gauge-fixing and 
Faddeev-Popov terms [13].

We have studied the temperature behavior of a strongly interacting 
Higgs system.
Using a gauged Chiral Lagrangian we have found that there are
no temperature dependent gauge corrections at one loop: the tendency 
towards symmetry
restoration observed in the ungauged case is then mantained. The 
computation of gauge corrections at two loops is still in progress.  
Although the results presented here are for a $SU(2)_L$ gauge symmetry, we are currently 
studying a theory with the Standard Model gauge symmetry, $SU(2)_{L} \times U(1)_{Y}$.
The model dependence, i.e. different symmetry breaking sectors of the theory 
(technicolor, SM...), is also being studied.  

It is a pleasure to thank the  Organizers of this Conference for the 
opportunity of presenting this work, and 
specially J.M. Fr\`ere for his help and patience. I am indebted with 
M.B. Gavela, M.J. Herrero, O. P\`ene and N. Rius for  many illuminating 
discussions and for 
crucial comments in the ellaboration of this manuscript. I thank 
A.M. Turiel and M.A. V\'azquez-Mozo for style corrections.

\section*{References}

\baselineskip=12pt
{\small
\noindent
\hspace{2pt} [1]  D.A. Kirzhnits, {\it JETP Lett.} {\bf 15} (1972), 529; D.A.
Kirzhnits, A.D. Linde, {\it Phys. 
\indent
\hspace{1pt} Lett.} {\bf B42} (1972) 47.

\noindent
\hspace{2pt} [2]  L. Dolan, R. Jackiw, {\it Phys. Rev.}, {\bf D9}, (1974) 3320.

\noindent
\hspace{2pt} [3] S. Weinberg, {\it Phys. Rev.} {\bf D9}, (1974) 3357.

\noindent
\hspace{2pt} [4]  R.N. Mohapatra, G. Senjanovic, 
{\it Phys. Rev. Lett.} {\bf 42}, (1979) 1651; 
{\it Phys. Rev.} 
\indent
\hspace{1pt} {\bf D20} (1979) 3390.

\noindent
\hspace{2pt} [5]  G. Amelino-Camelia, hep-ph/9610262; J. Orloff, 
hep-ph/9611398.

\noindent
\hspace{2pt} [6] G. Dvali, G. Senjanovic, {\it Phys. Rev. Lett.} {\bf 74} 
(1995); G. Dvali, A. Melfo, 

\indent
\hspace{1pt} G. Senjanovic, {\it Phys. Rev. Lett.} {\bf 75}
(1995), 4559 and references therein.

\noindent
\hspace{2pt} [7] H. Haber, {\it Phys. Rev.} {\bf D46} (1982), 1317;  M.
Mangano, {\it Phys. Lett.} {\bf 147B} (1984), 
\indent
\hspace{1pt} 307.

\noindent
\hspace{2pt} [8]  A. Riotto, G. Senjanovic, hep-ph/9702319.

\noindent
\hspace{2pt} [9] } G. Dvali, K. Tamvakis, hep-ph/9602336;  B. Bajc,
 A. Melfo, G. Senjanovic, 
\indent
\hspace{1pt} hep-ph/9607242.

\noindent
[10] A.T. Davies, C.D. Froggatt, G. Jenkins, R.G. Moorhouse, {\it Phys. 
Lett.} 

\indent
\hspace{1pt} {\bf B336} (1994), 464.

\noindent
[11] S. Weinberg, {\it Physica} {\bf A96} (1979), 327. 

\noindent
[12] J. Gasser, H. Leutwyler, {\it Phys. Lett.} {\bf B184} (1987), 83;
P. Gerber, 

\indent
\hspace{1pt} H. Leutwyler, {\it Nucl. Phys.} {\bf B321} (1989), 387;
A. Schenk, {\it Phys. Rev.} {\bf D47} 

\indent
\hspace{1pt}              
(1993), 5138; A. Bochkarev, J. 
Kapusta, {\it Phys. Rev.} {\bf D54} (1996), 4066. 

\noindent
[13]  T. Appelquist, C. Bernard, {\it Phys. Rev} {\bf D22} (1980), 20 
A. Longhitano, {\it Nuc. 
\indent
\hspace{1pt}
Phys.} {\bf B188} (1981), 118.

\end{document}